# The doping dependence of T* - what is the real high-$T_c$ phase diagram?

J. L. Tallon[a] and J.W. Loram[b]

[a] Industrial Research Ltd., P.O. Box 31310, Lower Hutt, New Zealand (J.Tallon@irl.cri.nz).
[b] IRC in Superconductivity, Cambridge University, Cambridge CB3 0HE, England

**Abstract**

Underdoped high-$T_c$ superconductors are frequently characterised by a temperature, T*, below which the normal-state pseudogap opens. Two different "phase diagrams" based on the doping (p) dependence of T* are currently considered: one where T* falls to zero at a critical doping state and the other where T* merges with $T_c$ in the overdoped region. By examining the temperature dependence of the NMR Knight shift and relaxation rate, entropy, resistivity, infrared conductivity, Raman scattering, ARPES and tunnelling data it is concluded that the second scenario is not at all supported. Neither can one distinguish a small and a large pseudogap as is often done. T* is an energy scale which falls abruptly to zero at p=0.19.

## 1. Introduction

The underdoped region of the high-$T_c$ superconductors (HTS) is distinguished by the presence of normal-state correlations, referred to as the pseudogap, which set in below a characteristic temperature, T* [1]. Discovered first as a systematic reduction on cooling in the $^{89}$Y NMR Knight shift by Alloul *et al.* [2] and in the spin-lattice relaxation rate $1/^{63}T_1$ by Yasuoka *et al.* [3], the pseudogap has a profound effect on a wide range of other physical properties and, indeed, is generally considered to be intimately connected with the origins of high-$T_c$ superconductivity. However, in spite of the many models that have been advanced [1] there is currently no consensus as to its nature. The present state of confusion regarding the pseudogap can be attributed in no small part to the fact that the HTS community cannot even agree on the detailed generic doping dependence of T*. After 13 years of intense study we are still left with the question: what is the real "phase diagram"? The two scenarios generally contemplated are summarised in Fig. 1. In (a) T* falls from a high value at low doping, comparable to the exchange energy J (≈1200K), and falls abruptly to zero at a critical doping point [4], while in (b) T*(p) merges with $T_c$(p) on the overdoped side [5]. In another variation of scenario (b) T*(p) is portrayed as rising only as high as the Neel temperature, $T_N$, as p→0 [6]. The distinction between these two scenarios is not trivial for, in the former case, the pseudogap is necessarily independent and competing with superconductivity and could, for example, implicate a quantum critical point [7], while the latter suggests that the pseudogap is some form of precursor pairing which evolves into the phase-coherent SC state at $T_c$ [5]. Scenario (b) is also implicated by strong-correlation spin-charge separation models [8] which associate T* with the formation of a gap in the spinon spectrum which is itself an essential precursor to the condensation of both spin and charge degrees of freedom below $T_c$.

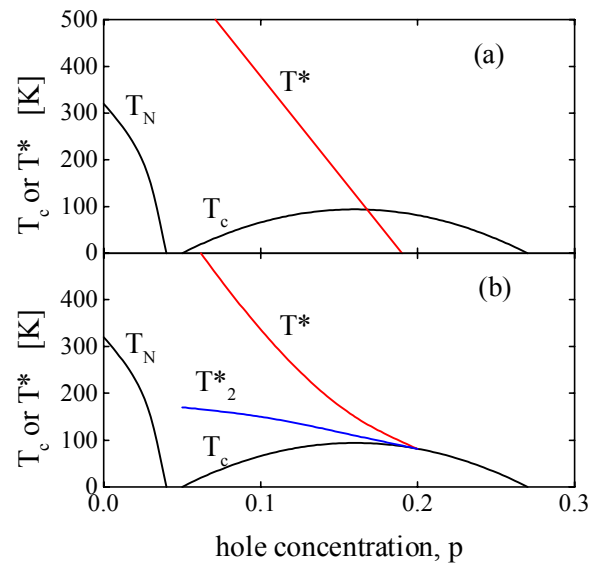

Fig. 1. Two scenarios for the "phase diagram" for HTS cuprates. In (a) T* represents and energy scale which falls abruptly to zero at a critical doping, p=0.19. In (b) T* merges with $T_c$ on the overdoped side and often a lower $T^*_2$ associated with a small pseudogap or a spin gap is invoked. $T_N$ is the Neel temperature for the 3-D AF state.

If either "phase diagram" could be conclusively demonstrated then a number of current models of high-$T_c$ superconductivity would have to be reviewed or possibly even abandoned. Here, by examining a wide range of physical properties, it is demonstrated that the actual scenario is that summarised by (a), namely that T*(p) falls definitely and abruptly to zero at a critical and *universal* doping state lying in the slightly overdoped region. Though the actual value of this critical doping concentration is of less significance and may continue to be debated, we identify it at p=0.19 [4,7] while optimal doping occurs at p=0.16. Moreover, by considering a variety of physical properties we show why it is that scenario (b) has been so often erroneously concluded. No consideration is given to the possibility of phase separation in the cuprate phase behaviour e.g. due to proximity to a van Hove singularity [9]. While there is nothing in our results that explicitly excludes this, we note that band-structure calculations result in singularities that are very system-dependent while the behaviour we discuss appears to be quantitatively universal amongst the cuprates.

**2. The Phenomenology of T***

T* is typically estimated by many different techniques including: the downturn in resistivity from linear behaviour above T*; the downturn in Knight shift or static susceptibility from the flat Pauli-like susceptibility observed at optimal doping; the position of the peak in $1/^{63}T_1T$; the temperature above $T_c$ at which the leading-edge gap in ARPES measurements falls to zero or at which the zero-bias tunnelling conductance fills out to the large-bias value; the temperature at which spectral weight in the far infrared c-axis conductivity begins to fall; and, the temperature at which the $2\Delta$ peak in Raman spectroscopy first starts to appear. Comprehensive references for these effects may be found in the review by Timusk and Statt [1]. The notion of a "phase diagram" delineated by a p-dependent T* emerged fitfully from the strong correlation theory of Nagaosa and Lee [8], and transport studies by Bucher *et al.* [10] on single crystals of $YBa_2Cu_4O_8$ and subsequently by Ito *et al.* [11] on $YBa_2Cu_3O_{7-\delta}$ and Batlogg *et al.* [12] based on scaling of the Hall coefficient in $La_{2-x}Sr_xCuO_4$ [13]. In each of these experimental studies the changes in the T-dependence of the resistivity and Hall coefficient were linked to systematic changes in Knight shift and spin-lattice relaxation rate associated with the opening of a spin gap or a pseudogap. While the term "spin gap" is still widely used, Loram *et al.* [14] showed from differential heat capacity studies that the pseudogap was consistent with a gap in the quasiparticle spectrum (and not just in the spin spectrum), a fact which was subsequently confirmed by ARPES [15] and tunnelling [16] studies. T* is widely viewed as a crossover temperature at which the pseudogap opens. It lies well above $T_c$ in underdoped samples, decreases rapidly with increasing doping and, in sufficiently overdoped samples, no T* feature is generally observable thus logically indicating that T*≤$T_c$. It is fair to say that in many instances experimentalists have unjustifiably adopted the upper limit T*=$T_c$, probably on ideological rather than logical grounds. An added problem is that most of these studies have been conducted for widely-separated doping states, typically for three states only: underdoped, overdoped and approximately optimally doped. Often little effort is expended in precisely determining optimal doping yet we know that, for example, very large changes in condensation energy and superfluid density occur around optimal doping with little change in $T_c$ [17,18]. A 90K Y-123 or Bi-2212 sample may be a little underdoped or a little overdoped but with widely differing properties depending upon which. Those few experiments [2,11,12,18,19] which utilise many small systematic increments of oxygen content and hole concentration have allowed the detailed evolution of T* with doping to be determined with a great deal more precision than other studies. Additionally and to be blunt, the subjective identification of T* from a perceived downturn in noisy data is at times demonstrably naïve. The visual perception of a downturn in any case depends upon just how the data is plotted (as will be shown when we consider the resistivity and the spin-lattice relaxation rate) and, more importantly, we demonstrate that the temperature evolution of most properties is in fact very smooth across T* with no identifiable abrupt change that might suggest e.g. an underlying phase transition to a long-range ordered state. The smoothly evolving behaviour is only consistent with an underlying energy scale and T* must be considered as an energy not a temperature. For this reason we refer to the term "phase diagram" in quotation marks for there is no distinct pseudogap phase in the thermodynamic sense.

Many authors deduce a second lower crossover temperature $T^*_2$ as shown in Fig. 1(b). According to various reports of this view (e.g. Schmalian *et al.* [20]) $T^*_2$ is the small T* obtained from $1/^{63}T_1T$, the resistivity and the leading-edge ARPES gap, while the Knight shift, static susceptibility and the ARPES "dip-hump" feature reveal the large T* crossover. Tunnelling and heat capacity data have been variously placed on either line. Needless to say there is a good deal of scatter when all this is put together. These difficulties, we suggest, are conceptual and interpretive and, for example, do not arise from variations in the determination of the hole concentration since a large number of authors now relate p to $T_c$ using the approximately parabolic relation of Presland *et al.* [21]. In the following we separately consider in detail each of these physical properties.

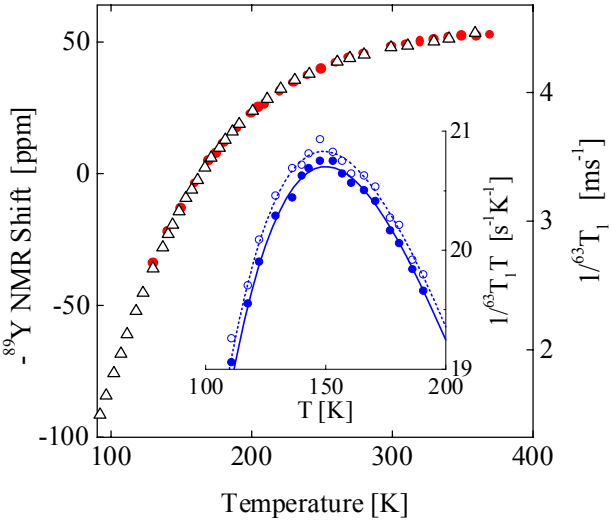

Fig. 2. $^{89}$Y NMR magic-angle spinning Knight shift (●) and copper relaxation rate (Δ) for Y-124. Inset: $1/^{63}T_1T$ for Y-124 with oxygen isotopes $^{16}$O (○) and $^{18}$O (●).

## 3. Physical properties

*3.1 Knight shift.*

The T-dependence of the $^{89}$Y NMR Knight shift for YBa$_2$Cu$_4$O$_8$ (Y-124) is shown in Fig. 2 (filled circles) [22]. This is amongst the most precise reported Knight shift data for any HTS cuprate due partly to the absence of defects in this compound as well as the use of precise temperature-control and magic-angle spinning. Line-widths were of the order of just 100Hz. Typically a T* value of 180K is assigned to Y-124 but it is clear that the data is absolutely free of any kink or discontinuity (even in higher derivatives) that could allow such an assignment. One can only say that there is an underlying energy scale and extract this by fitting, say, to a DOS with a triangular (or V-shaped) gap at the Fermi level with gap energy $E_g \approx 370$K. T* ($\approx 0.4 E_g/k_B$) must then be viewed as an energy scale. More generally, as shown by Wuyts *et al.* [19], the succession of Knight shift curves $^{89}K_s(T)$ obtained by Alloul *et al.* [2] for different doping states across the underdoped region may be scaled to a single function of $T/E_g$. Fig. 3 shows a similar scaling analysis of $^{89}K_s(T)$ for Y$_{0.8}$Ca$_{0.2}$Ba$_2$Cu$_3$O$_{7-\delta}$ [23]. The p-dependence of $E_g$ values determined in this way is plotted as triangles and circles in the inset to Fig. 3 and as circles in Fig. 4. These fall linearly to zero at the critical doping state of p=0.19 and rise to the magnitude of the exchange energy, J, as p→0. Notably, the overdoped $^{89}K_s(T)$ data has a different scaling behaviour and the scaling separatrix between the under- and overdoped regions is p=0.19, *not* optimal doping at p=0.16. This gives additional credence to the existence of a well-defined point at which $E_g \to 0$. In particular, the data cannot be construed as defining a temperature, T*, where the Knight shift suddenly decreases in magnitude and which merges with $T_c(p)$ in the overdoped region.

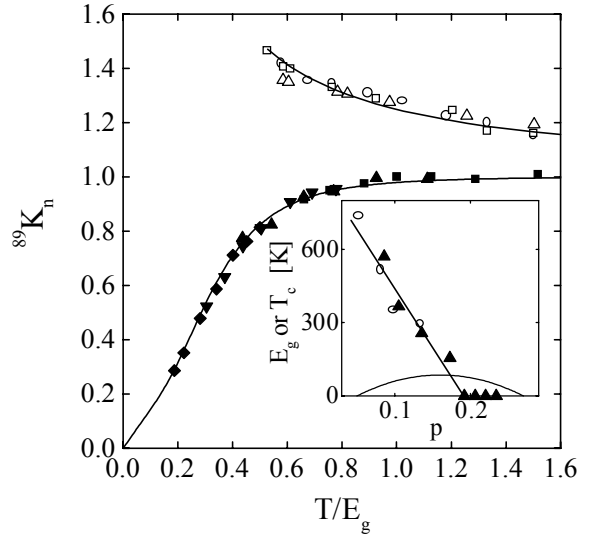

Fig. 3. The $^{89}$Y Knight shift for Y$_{0.8}$Ca$_{0.2}$Ba$_2$Cu$_3$O$_{7-\delta}$ with different δ values scaled as a function of $T/E_g$ (◆: $T_c$=47.5, p=0.086; ▼: $T_c$=65.8, p=0.107; ▲: $T_c$=83.2K, p=0.140; ■: $T_c$=86K, p=0.160; Δ: $T_c$=72.1K, p=0.204; □: $T_c$=60K, p=0.221; ○: $T_c$=47.5K, p=0.234). Insert: $E_g$ values obtained from the scaling, 0.1Ca (○) and 0.2Ca (▲).

*3.2 Heat capacity.*

Firstly, it is clear that T* or $E_g$ for the heat capacity and, in particular, the entropy, S, is the same *large* pseudogap as that found for the Knight shift and static susceptibility, $\chi_s$. Indeed, when multiplied by an appropriate Wilson ratio, S/T above $T_c$ is found to equal $\chi_s$ and $K_s$ over a broad range of T and p [24]. Previously reported values of $E_g$ obtained by a similar scaling analysis of S/T for Y$_{0.8}$Ca$_{0.2}$Ba$_2$Cu$_3$O$_{7-\delta}$ are shown by the crosses in Fig. 4. They agree very well with the values obtained from Knight shifts and, again, they reduce with increasing doping from the magnitude of J to fall to zero at $p=0.19$. Values of the condensation energy $U_o$ obtained by integrating the entropy data for each doping state [18] are also plotted in Fig. 4. The fact that they pass through a sharp peak at $p=0.19$ further underscores the unique and sudden character of this critical doping point. Note that the $E_g$ values are obtained from normal-state data but the $U_o$ values are T=0 properties reflecting the change in free energy in transforming from the normal state to the superconducting ground state. Both identify the same critical doping point. Values of the low-temperature superfluid density obtained from muon spin relaxation on the same samples also pass through a sharp peak at $p=0.19$ as do magnetisation critical currents for these samples at 10K [7]. Taken together, there is a remarkable concurrence between normal-state properties and the ground-state superconducting properties that there exists a unique critical doping point in the phase diagram where properties change very suddenly, where the pseudogap energy falls to zero and where superconductivity is most robust. These features are not consistent with the precursor-pairing scenario. Why would phase fluctuations, induced by a low superfluid density, set in just where the superfluid density is at a maximum?

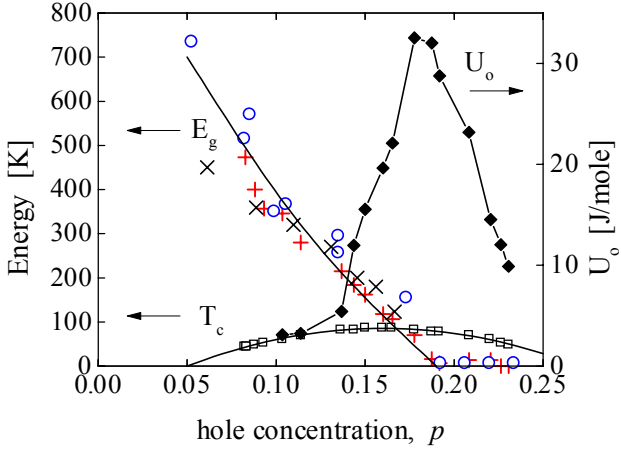

Fig. 4. The $p$-dependence of $E_g$ for $Y_{0.8}Ca_{0.2}Ba_2Cu_3O_{7-\delta}$ from $^{89}K_s$ (O), from heat capacity (+) and from scaling of: resistivity (×), condensation energy $U_o$ (◆) and $T_c$ (□).

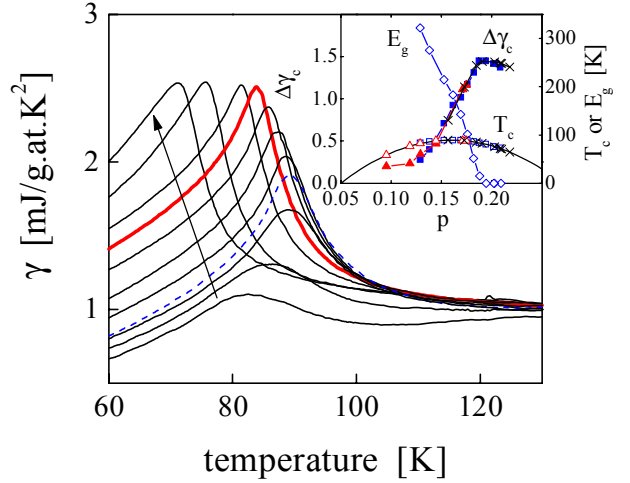

Fig. 5. The T-dependence of $\gamma \equiv C_p/T$ for Bi-2212 with different oxygen contents spanning from underdoped to overdoped (indicated by direction of arrow). The curves for critical and optimal doping are indicated by the bold and dashed curves, respectively. Inset: the doping dependence of the increment in $\gamma$ for $Bi_{2.1}Sr_{1.9}CaCu_2O_{8+\delta}$ (■), $Bi_{1.9}Pb_{0.2}Sr_{1.9}CaCu_2O_{8+\delta}$ (+) and $Bi_{2.1}Sr_{1.9}Ca_{0.7}Y_{0.3}Cu_2O_{8+\delta}$ (▲). In each case $\Delta\gamma_c$ falls abruptly at $p=0.19$ with the opening of the pseudogap. $E_g$ values obtained from a scaling analysis are shown by the diamonds.

An additional factor is that this behaviour of the heat capacity in Y:Ca-123 is common to other cuprates. Fig. 5 shows the T-dependence of $\gamma(T) \equiv C_p/T$ for a range of different oxygenation states in $Bi_2Sr_2CaCu_2O_{8+\delta}$ (Bi-2212) spanning either side of optimal doping [25]. The overdoped samples exhibit a jump, $\Delta\gamma_c$, in $\gamma$ at $T_c$ which remains more or less constant. Then beginning with the opening of the pseudogap at critical doping $\Delta\gamma_c$ falls sharply, and already by optimal doping has reduced to about 50% of its maximum value before further diminishing on the underdoped side. The raw data requires no further analysis in the form of scaling or modelling to convincingly demonstrate just how sudden is the onset of the pseudogap at critical doping right at the point where the rigidity of the condensate wave function is at its maximum. Such behaviour would be very difficult to reconcile with a phase fluctuation scenario but is very consistent with the onset of correlations which compete with superconductivity for the same section of Fermi surface, taking away spectral weight otherwise available for superconductivity [14]. It is also notable in Fig. 5 that there is no change with doping in the fluctuation contribution to $\gamma(T)$ in the neighbourhood of $T_c$. The fundamental change in fluctuations as $T^* \rightarrow T_c$ expected within a precursor pairing scenario is not seen in the data. The inset to Fig. 5 summarises the doping dependence of $\Delta\gamma_c$ and of $T_c$ for three different Bi-2212 samples: pure, 0.20Pb substituted and 0.15Y substituted samples which variously push deeper into the overdoped or underdoped regions, respectively, while in all cases spanning critical doping. In each case, irrespective of their different compositions and oxygen content, the critical doping point where $\Delta\gamma_c$ sharply falls is the same, namely $p=0.19$. Again, this is the point where $E_g \rightarrow 0$ (as shown in the insert) and $U_o$ maximises. This behaviour would appear to be generic in the HTS cuprates [18].

## 3.3 Spin-lattice relaxation rate.

The spin-lattice relaxation rate is the rate at which the nuclear magnetisation relaxes to thermal equilibrium and is given by

$$1/T_1T \sim \sum_{\mathbf{q}} |A_\mathbf{q}|^2 \chi''(\mathbf{q},\omega_o) / \omega_o. \qquad (1)$$

where $A_\mathbf{q}$ are the hyperfine coupling form factors, $\omega_o$ is the NMR or NQR frequency and $\chi''(\mathbf{q},\omega_o)$ is the imaginary part of the dynamical spin susceptibility. For the HTS cuprates the form factors in $1/^{63}T_1$ maximise at $\mathbf{q}=0$ and $\mathbf{q}=\mathbf{Q}\equiv(\pi,\pi)$ while $\chi''(\mathbf{q},\omega_o)$ maximises at $\mathbf{q}=\mathbf{Q}$. Therefore $1/^{63}T_1T$ is dominated by the response at $\mathbf{Q}=(\pi,\pi)$. Values of the relaxation rate are rather large due, as is generally believed, to scattering from antiferromagnetic (AF) spin fluctuations. As a consequence $1/^{63}T_1T$ adopts a Curie-like $1/T$ dependence at high temperature. At a lower temperature, $T^*_2$, this quantity passes through a maximum then falls rapidly as shown by the two curves for Y-124 in the inset to Fig. 2. These curves are for $^{16}O$ or $^{18}O$ isotope exchange as reported by Raffa *et al.* [26]. The rapid fall below $T^*_2$ was, and to some degree is still, widely viewed as due to the opening of a spin gap but more recently variously attributed to the onset of a "strong pseudogap" [20] or of precursor pairing [5]. It is clear, however, that $T^*_2$ does not represent any such onset. The isotope effect in $1/^{63}T_1T$, shown in the inset, does not set in at $T^*_2$ but is uniformly present across the entire T-range. Indeed, if the temperature dependence of $1/^{63}T_1$ (without the $1/T$ enhancement) is examined it is found to be identical to that for $^{89}K_s$ for Y-124 (see triangles in Fig. 2, main panel). This can be understood within the enhanced susceptibility of Millis-Monien-Pines (MMP) [27] which, when substituted into equ. (1) in the limit $\xi_{AF}^2 \gg 1$, yields [28,29]

$$1/^{63}T_1 = a_1 \chi_s T/\omega_{SF} . \qquad (2)$$

Here $a_1$ is a constant and $\omega_{SF}$ is the spin-fluctuation paramagnon frequency. From this relation the proportionality shown in Fig. 2 naturally arises if $\omega_{SF} \propto T$ for $T > 100K$ and would appear to be evidence of such proportionality. This result, that $1/^{63}T_1 \propto {}^{89}K_s$, shows that the spin-lattice relaxation rate and the Knight shift for Y-124 have *exactly* the same underlying energy scale given by $E_g$ and, in particular, that the drop in $1/^{63}T_1T$ does not define an onset of any sort, whether a spin gap or precursor pairing. It simply reflects the same large pseudogap seen in $\chi_s$ and S/T. Conversely, if one were to plot the T-dependence of $\chi_s/T$ it is apparent that this would peak at the same $T^*_2$ value as $1/^{63}T_1T$ but, of course, this has no physical significance in terms of an onset.

As noted, the Curie-like T-dependence of $1/^{63}T_1T$ in the high-temperature limit is viewed as a signature of AF correlations. In particular, La-214 exhibits Curie-like behaviour across the overdoped region [30] thus suggesting the persistence of spin fluctuations throughout this region. However, reference to equ. (1) shows that this can be misleading. For overdoped La-214, $\chi_s$ exhibits a strong and growing upturn at lower temperatures [31] that is not present to the same degree in other HTS cuprates. This clearly makes the inference of AF correlations from the T-dependence of $1/^{63}T_1T$, alone, more ambiguous. A better route is to consider the ratio $^{17}T_1/^{63}T_1$.

The spin-lattice relaxation rate, $1/^{17}T_1$, for $^{17}O$ may be found by substituting the MMP enhanced susceptibility in equ. (1). When $\xi^2 \gg 1$ this yields:

$$1/^{17}T_1T = a_2 \chi_s . \qquad (3)$$

This appears to be well satisfied for a number of HTS cuprates [28] and implies that the pseudogap energy scale, $E_g$ is also to be found in $1/^{17}T_1T$ [28]. The merit of considering the ratio $^{17}T_1/^{63}T_1$ is that $\chi_s$ is divided out. More generally,

$$^{17}T_1/^{63}T_1 = a_4 <1 + f_\mathbf{q}^2> = a_4[1+C_{AF}/T] \qquad (4)$$

where $f_\mathbf{q}$ is the ratio of the MMP enhanced susceptibility near $\mathbf{Q}=(\pi,\pi)$ to its weakly-interacting Fermi liquid value and the average is over $\mathbf{q}$. $C_{AF}$ thus measures the strength of AF correlations in the normal state. The value of $C_{AF}$ has been extracted by fitting [32] the data for the T-dependence of $^{17}T_1/^{63}T_1$ for Y-123, Y-124 and oxygen-deficient Y-123 and is plotted in Fig. 6 as a function of hole concentration (solid diamonds). This can be seen to fall steadily towards zero at critical doping indicating that the pseudogap is intimately associated with short-range AF correlations which disappear abruptly at critical doping (at least on the NMR time scale of $10^{-7}$ s). As we will see, other probes present a similar picture.

## 3.4 Resistivity.

The small pseudogap, or $T^*_2$, reported from the resistivity is typically deduced by visually locating a point where $\rho(T)$ departs from linearity. $T^*_2$ is reported to merge with $T_c$ on the overdoped side. Such a construction is ill-conceived. As pointed out by Hussey *et al.* [33], $(\rho(T) - \rho_o)/aT$ for high-quality single crystals of Y-124 exhibits a smooth rise over the entire temperature range from $T_c$ to 400K and is absolutely fea-tureless at $T^*_2 = 180K$, as are higher derivatives. Here $\rho_o$ and $a$ are the intercept and slope, respectively, of the linear T-dependence seen at high temperature. Moreover, Ito et al. [11] have shown that $(\rho(T) - \rho_o)/aT$ has the same doping and temperature dependence as the Knight shift. Elsewhere [28], it was shown that, for Y-124, $d\rho/dT$ is precisely proportional to $d^{89}K_s/dT$ from 120K to 400K. These two results indicate that $\rho(T)$ has the same underlying large energy scale as the spin susceptibility. ($\rho(T)$ must contain a similar Fermi-window average of the DOS to that contained in $\chi_s$). Given the scaling behaviour of $\chi_s(T)$ it is therefore not surprising that $\rho(T)$ also scales to a single function of $T/T^*$. The original attempts to scale the resistivity, however, were motivated by the reported scaling of the Hall coefficient by Hwang et al. [13] for La-214 and later by Chen et al. [34] for Y-123 films. Such a scaling analysis of the resistivity has been carried out by Wuyts et al. [19] for single-crystal Y-123, by Tallon et al. for thin films of $Y_{0.8}Ca_{0.2}Ba_2Cu_3O_{7-\delta}$ [7] and by Konstantinovic et al. [35] for thin films of Bi-2201 and Bi-2212. These all show that the energy scale $T^*\to 0$ at critical doping (the crosses (×) in Fig. 4 are the data of Wuyts et al. [19]) and can in no way be construed as $T^*\to T_c$. Interestingly, critical doping is the only point at which $\rho(T)$ is linear from $T_c$ to the highest temperatures [7], rather suggestive of the expected behaviour above a quantum critical point. If the hole concentration is reduced a little from critical doping to optimal doping significant curvature, associated with the opening of the pseudogap, sets in from about 200K [7].

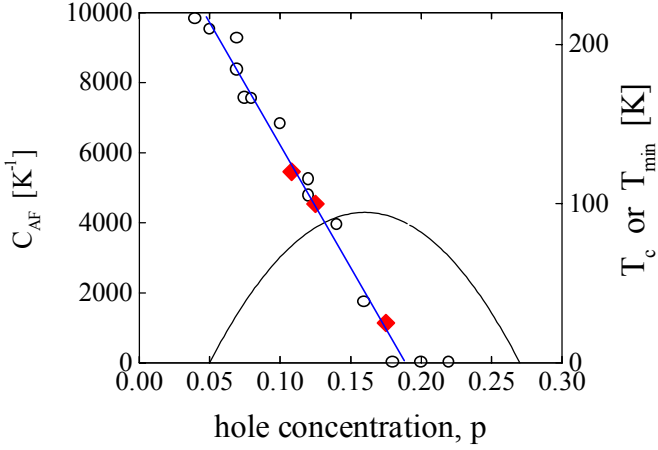

Fig. 6. The doping dependence of the position, $T_{min}$, of the minimum in a-b plane resistivity for La-214 (○) and of the amplitude $C_{AF}$ of short-range AF correlations (◆) determined from the ratio $^{17}T_1/^{63}T_1$ using equ. (4).

If there is any doubt in this matter it is instructive to simulate $\rho(T)$ and carry out the usual identification of $T^*$. Suppose, $\rho(T)$ is linear in T for $T\gg T^*$ and that the effect of the pseudogap is modelled like the susceptibility as $[1 - \theta^{-1} \tanh\theta \ln(\cosh\theta)]$ where $\theta \equiv E_g/k_B T$. This function represents a triangular (nodal) gap that fills with increasing temperature and quite accurately characterises the T-dependence of the Knight shift and the electronic entropy. Thus

$$\rho(T) = (a+bT) \times [1 - \theta^{-1} \tanh\theta \ln(\cosh\theta)]. \quad (5)$$

In addition, the superconducting transition and the paraconductivity above $T_c$ are modelled by assuming 2D fluctuations where the fluctuation conductivity $\Delta\kappa$ is given by [36]:

$$\Delta\kappa = (\hbar/16e^2 d)(T/T_c - 1)^{-1}. \quad (6)$$

where d is the mean spacing between $CuO_2$ layers. The p-dependence of $E_g$ is taken from the linear fit shown in the inset to Fig. 3 and the p-dependence of $T_c$ is given by the parabolic approximation [21]

$$T_c = T_{c,max} [1 - 82.6 (p-0.16)^2]. \quad (7)$$

Fig. 7 shows the calculated resistivity for p=0.13 (with $E_g$=284K) plotted to 300K in (a) and to 600K in (b). By fitting the visibly linear portion from 600K down a clearly visible downturn (arrowed) is revealed at $T^*\approx 320K$. When, however, the function is plotted only to 300K, fitting the apparently linear portion from 300K down gives a discernible break at a much lower temperature $T^*\approx 195K$. The large difference between these two values obtained from the same set of data shows that the visual method is clearly too subjective to be quantitative. Moreover, the simulation shows that such problems are further exacerbated when $T^*$ is close to $T_c$ as shown in the inset to Fig. 7(a) for the case p = 0.17 ($E_g$=95K). The eye cannot judge between superconducting fluctuations and a pseudogap-induced downturn. The result is the false impression that $T^*$ merges with $T_c$.

This is precisely the case with fully oxygenated Y-123 which is slightly overdoped and close to critical doping. The downturn (due to superconducting fluctuations) which sets in below 120K could inadvertently be interpreted as the opening of the pseudogap with $T^* \approx 120K$. Progressive Zn substitution in Y-123 films reduces both $T_c$ and the downturn which, for all cases, sits ~ 20K above $T_c$ [37] confirming that its origin is indeed merely superconducting fluctuations. In the 7% Zn substituted sample $T_c$ is as low as 20K and $\rho(T)$ is featureless above 35K [37], showing no trace of the original downturn at 120K and confirming that $T^* < 35K$ (and probably zero). In this context it should be noted that Zn substitution has no effect on a genuine T* downturn in underdoped samples [38]. In summary, the only proper method for analysis of ρ(T) is a scaling analysis which, as noted, yields T* → 0 as p → 0.19.

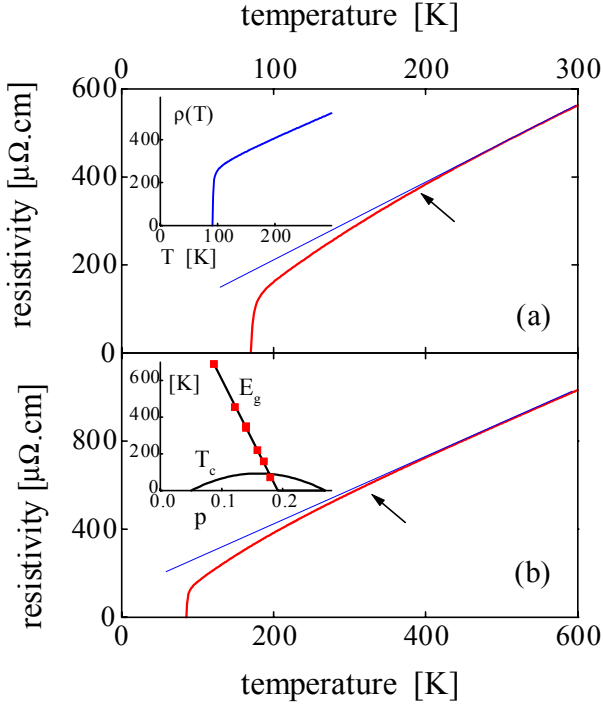

Fig. 7. The simulated resistivity for p=0.15, following equ. (5). (a) When plotted on a scale of 0 to 300K T* (arrowed) appears to be at 195K but (b) when plotted 0 to 600K T* ≈ 320K. Inset (a): the simulated resistivity for p=0.17 where T* is close to $T_c$. Inset (b): values of $E_g$ determined from fitting the c-axis resistivity (from ref. [41] ).

We conclude then, that, as for the Knight shift, entropy and relaxation rate, the resistivity energy scale is set by the large pseudogap which vanishes at critical doping. The resistivity features discussed above appear to be dominated by effects on scattering rate which lower the resistivity. Pseudogap DOS effects on the carrier density (which raise the resistivity) do not seem to appear until lower temperatures with the crossover from metallic to insulating behaviour where the resistivity diverges as log(-$T/T_{min}$) [39]. The crossover is conveniently demarked at $T_{min}$ by the minimum in ρ(T). The doping dependence of $T_{min}$ for La-214, where the superconductivity is suppressed either by intense magnetic fields or by Zn substitution, has been investigated elsewhere [32] and is reproduced in Fig. 6 by the open circles. $T_{min}$ is seen to fall linearly to zero at p=0.19. Critical doping is thus not only the point at which the normal-state pseudogap energy falls to zero and where superconductivity is most robust but the point where the normal-state metal-insulator transition occurs at T=0 in the absence of superconductivity.

*3.5 c-axis resistivity.*

The metal/insulator transition at critical doping is also observed in the c-axis resistivity, $ρ_c ≡ 1/σ_c$, which crosses over from a semiconducting to metallic T-dependence apparently *independent of the anisotropy*. This crossover has been attributed to the closing of the pseudogap [40] and more recently modelled by Cooper and Loram [41] in terms of a simple Giaever tunnelling model using

$$σ_c = (2π/\hbar)\; e^2 c^2 t_c^2 \int N(E)^2 × ∂f/∂E\; dE , \qquad (8)$$

where f(E) is the Fermi function, c is the c-axis lattice constant and $t_c$ is the interplanar tunnelling matrix element which is assumed, as in usual superconducting tunnelling, to be independent of the initial and final momenta. Using a triangular density of states N(E) the $ρ_c$ data for single crystal $YBa_2Cu_3O_{7-δ}$ with different δ values was accurately fitted to obtain values of the gap energy $E_g$ [41]. The δ values have been converted to values of hole concentration and these $E_g$ values are plotted as a function of p in the inset to Fig. 7(b). Again, values of $E_g$ are remarkably consistent with those determined using other techniques and they show a clear sudden closure of the pseudogap at critical doping $p_{crit} ≈ 0.19$.

*3.6 Infrared conductivity*

The crossover from insulating to metallic behaviour in the c-axis resistivity is also reflected in the *sudden* appearance of a Drude peak in the c-axis frequency-dependent normal-state conductivity, $σ_c(ω)$, reported in a series of studies on $Y_{1-x}Ca_xBa_2Cu_3O_{7-δ}$ [42]. In the case of $La_{2-x}Sr_xCuO_4$ the Drude peak in $σ_c(ω)$ can be seen from the data of Henn *et al.* [43] to appear suddenly between p=x=0.18 and 0.20. This behaviour is again confirmed in an independent study by Uchida and coworkers [44]. It is difficult to avoid the conclusion that in the HTS cuprates there is a universal crossover at critical doping to coherent c-axis transport, irrespective of anisotropy, which is governed solely by the doping state. This circumstance is consistent with the fact that c-axis transport is dominated by parts of the Fermi surface near the zone boundary at (π,0) [45-47]. In underdoped samples quasiparticles near these points are strongly damped due to scattering from AF spin fluctuations with the result that c-axis transport is not coherent. The *sudden* recovery of the Drude peak in $σ_c(ω)$ at critical doping signals the recovery of long quasiparticle lifetimes near ($\pi$,0) and is indirect evidence that the short-range AF background becomes abruptly suppressed at this doping point. This observation will frequently re-emerge as a sub-theme in the following discussion of other physical properties of the cuprates.

It is instructive to examine the Ca-substituted Y-123 infrared conductivity data [42] in more detail. For the overdoped samples, beyond critical doping, the onset of superconductivity is characterised by a suppression of spectral weight below a frequency given by $2\Delta$, where $\Delta$ is the superconducting gap energy. The spectral weight is transferred into the conductivity delta function at $\omega=0$. $\Delta(T)$ is found to collapse to zero as $T\rightarrow T_c$, that is, the superconducting gap *closes* with increasing temperature [42]. In contrast, for underdoped samples spectral weight is suppressed beginning well above $T_c$, the weight is transferred to higher frequencies and the gap energy is approximately T-independent. The doping dependence of the size of the deduced low-temperature gap is plotted in Fig. 8 (open circles) and compared with the gap determined from SIS tunnelling studies [48] (solid squares). Both show a steady rise to large values with progressive underdoping. Also plotted are values of $4k_BT_c$. Naively, the data could be construed to indicate that the gap values behave like the doping dependence of a T* line which merges with $T_c$ on the overdoped side. The reality is rather different. The large gaps plotted for the underdoped region are those which appear at quite high temperatures (i.e. the pseudogap) and do not correspond to the superconducting gap. Indeed, the inset to Fig. 8 reproduces the T-dependent spectra for the most underdoped sample with $T_c$=78 K. One may clearly discern the pseudogap at 800 cm$^{-1}$ ($E_g$ = 575 K) which becomes apparent by the reduction in $\sigma_1(\omega)$ starting at about 200 K, and an independent and distinct superconducting $2\Delta_o$ gap at 500 cm$^{-1}$ ($\Delta_o$ = 360 K) which is established at $T_c$=78 K. In principle, the pseudogap will continue to further suppress spectral weight below 800 cm$^{-1}$ as the temperature is reduced below $T_c$ but the sudden additional suppression below 500 cm$^{-1}$ associated with the superconducting gap is clearly evident. While this data may be complicated by the presence of a possible Josephson plasma bilayer resonance at 410 cm$^{-1}$ [49,50] a separate SC gap can also be discerned in the T dependence of $\sigma(\omega)$ for single-layer cuprates [51].

Thus plots, such as shown in Fig. 8, of the p-dependence of the dominant gap in fact combine two separate gaps, $E_g$ and $\Delta_o$ and, for any doping state, it is the larger of the two gaps which is typically inferred. The two left-most IR gaps in the Fig. 8 are the pseudogap while the three right-most gaps are the super-conducting gap. This is quite clear from the conductivity data which shows no suppression of spectral weight above $T_c$ for the latter three samples. This interpretation is given additional credence by the values of the superconducting spectral gap, $\Delta_o$, obtained from heat capacity measurements [18]. The p-dependence of $\Delta_o$ is shown by the solid triangles in Fig. 8 and these can be seen to map onto the low-temperature IR and tunnelling gaps once $E_g$ falls below $\Delta_o$. These conclusions will be further substantiated by the tunnelling data considered in more detail below.

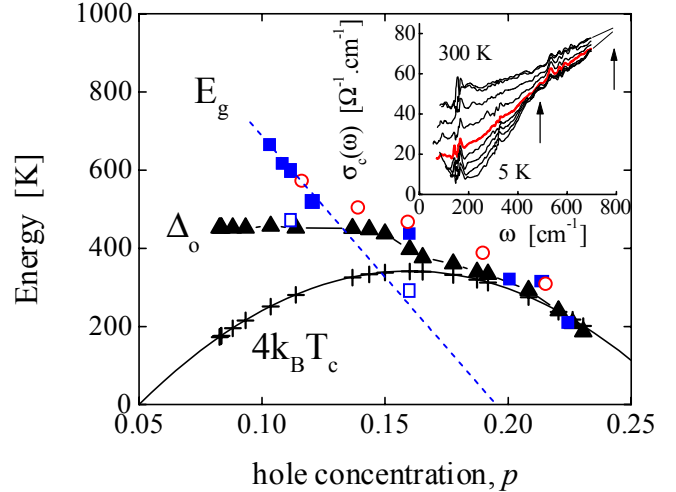

Fig. 8. p-dependence of the low-T spectral gap determined from c-axis IR spectroscopy for Y-123 (○) and SIS tunnelling for Bi-2212 (■) and $4k_BT_c$ for $Y_{0.8}Ca_{0.2}Ba_2Cu_3O_{7-\delta}$ (crosses). Also shown is the SC gap $\Delta_o$ determined for $Y_{0.8}Ca_{0.2}Ba_2Cu_3O_{7-\delta}$ from heat capacity (▲). Open squares are the additional second-gap tunnelling features that show that the low-doping gap is the pseudogap while the high-doping gap is the SC gap. Inset: the IR conductivity $\sigma_1(\omega)$ for an underdoped sample of Y-123 with $T_c$= 78 K see ref. [42]) for different temperatures (300, 200, 150, 100, 80, 70, 55, 450 and 5 K). The highlighted curve (80 K) is close to $T_c$. Arrows mark the distinct SC gap and pseudogap.

In summary, we see in the IR data the same general features originally indicated by the heat capacity and NMR data, namely, two independent and competing gaps where $\Delta(T)$ has a BCS-like mean-field T-dependence while $E_g$ is largely T-independent. The two gaps have quite distinct doping dependences and by optimal doping the pseudogap has diminished to less than 150 cm$^{-1}$ (107 K) [42].

*3.7 Raman spectroscopy.*

Similar conclusions also apply to measurements of the spectral gap observed in Raman spectroscopy. The p-dependence of the $2\Delta$ pairbreaking peak seen in $B_{1g}$ Raman measurements on Bi-2212 [52] actually tracks closely with the $2\Delta$ values determined from heat capacity which are shown in Fig. 8. However care must be exercised in interpreting this data in the underdoped region. The $B_{1g}$ channel probes the scattering response near the $(\pi,0)$ zone boundary. The pseudogap is principally manifested by a strong depletion of integrated spectral weight below 800 cm$^{-1}$ due to scattering from short-range AF spin fluctuations [53]. The weight is transferred to the region of the two-magnon scattering peak (with energy scale 3J) which itself is associated with the short-range AF background. The two-magnon scattering disappears abruptly at p=0.19 [54] confirming in our view the sudden loss of short-range AF correlations at critical doping [32]. Only with the near disappearance of the pseudogap does there occur a $2\Delta$ renormalisation which sets in at $T_c$, presumably because the $B_{1g}$ symmetry probes that part of the Fermi surface where the quasiparticles are strongly damped. In the more underdoped region the $2\Delta$ feature moves towards, then overlaps, the 600 cm$^{-1}$ peak incorrectly attributed by Blumberg *et al.* [55] to a bound precursor pairing state. This scattering peak has been suggested by Hewitt *et al.* [56] to be a phonon mode associated with oxygen disorder in the deoxygenated BiO layers. In support, they note that the scattering peak exhibits an oxygen isotope effect and is absent in oxygenated samples which are underdoped by Y substitution. Bearing in mind these additional subtleties, it is apparent that the $2\Delta$ feature in the Raman $B_{1g}$ symmetry is identically the same in magnitude and doping dependence as the superconducting gap observed in heat capacity. For the sake of clarity, these Raman gap energies have been omitted from Fig. 8.

The strong depletion of spectral weight associated with the pseudogap finds a direct analogue with the entropy depletion observed in the heat capacity measurements. In underdoped samples these lost states are not recovered in the 'high-temperature' (300K) entropy presumably because 300K is still small compared to the two-magnon energy scale (3J). The important point to take from the Raman studies in the present context is that, in the strongly underdoped samples, the pseudogap energy scale is set by J (much higher than $\Delta$) and appears to be rather temperature independent [53]. These have emerged as common features in all of the other spectroscopies discussed here. They are certainly inconsistent with a precursor pairing scenario and do not support a picture in which $T^* \to T_c$. Rather, they give further credence to a scenario in which the pseudogap is associated with short-range AF correlations which disappear abruptly at critical doping.

*3.8 ARPES and tunnelling.*

Each of the above techniques are bulk measurements. ARPES and certain tunnelling studies involve the serious uncertainties associated with shallow-surface spectroscopy. Much of this work has been carried out almost exclusively on Bi-2212. It is not at all clear how the surface properties, including the incommensurate superstructure together with excess oxygen and bismuth, are modified on cleavage *in vacuo*. In spite of these potential limitations, both of these techniques confirmed what had earlier been clear from the heat capacity [57, 24], namely that the pseudogap is a quasiparticle gap. Both techniques show gap features in underdoped samples persisting well above $T_c$. A simple analysis of the leading-edge ARPES gap shows that this progressively narrows and closes at a $T^*$ that decreases with doping [15] and is comparable in magnitude to some other measures of $T^*$. This would appear to directly counter the inference from $K_s$ and S/T that the gap remains rather T-independent above $T_c$ and fills rather than closes. However, such a leading-edge analysis inevitably yields an apparently T-dependent gap if there is any T-dependent broadening of the edge. Tunnelling studies, consistent with $K_s$ and $\chi_s$, more clearly show the slow progressive filling, rather than closing, of the gap [16].

A more critical issue is the fact that both ARPES and tunnelling have tended to suggest that the pseudogap evolves smoothly into the superconducting gap either with decreasing T [15,16] or with increasing p. This is illustrated in Fig. 8 where the SIS tunnelling gaps found by Miyakawa et al. [48] are plotted as a function of hole concentration (solid squares) along with the IR gaps discussed above (open circles). The variation with p is roughly like that of various measures of $T^*(p)$ in which $T^*(p) \to T_c(p)$ on the overdoped side. This was construed as strong evidence in favour of precursor pairing models of the pseudogap [48]. In fact, this apparent smooth evolution of the gap may be resolved in a manner fully consistent with an independent and competing pseudogap.

The STM technique has recently been developed to the point that, using a suitably stiff system and spatially stable tip, tunnelling over a single atom can be sustained for extended periods of time. This allows, for example, tunnelling over a single zinc substitutional impurity [58]. These studies show that the fresh-cleaved surface of Bi-2212 is a patchwork of locally-varying doping states so that some areas, extending over a few atoms, reveal sharp DOS features associated with the superconducting gap while other areas show larger gaps without coherence peaks, reminiscent of the pseudogap. Previous STM studies, due to tip instability, appear to have merely shown averaged features with broadened coherence peaks. It is therefore not surprising that the larger pseudogaps appeared to evolve smoothly into superconducting gaps. These broadened spectra conceal, though not completely [59], the presence of two distinct gaps. By focussing just on the low-temperature gap, in such broadened spectra one obtains only the largest gap and one sees a simple crossover with increasing doping from one gap to the other as illustrated in Fig. 8. On the other hand, the higher temperature spectra reveal additional discernible gap features [48] which are plotted as open squares in Fig. 8. The figure now reveals one gap that tracks the doping dependence of the superconducting gap and another that tracks the pseudogap as indicated by the dashed line. These features, on their own, might be considered too subtle to warrant such a conclusion. However, we note that recent intrinsic tunnelling studies on Bi-2212 thin-film mesas, which avoid the highly local nature of the STM technique, unequivocally reveal the two gaps with a sharp states-conserving superconducting gap and a broad non-states-conserving pseudogap [60]. The pseudogap energy, $E_g$, appears largely T-independent above and below $T_c$ consistent with the above-noted results from S/T and $K_s$, while $\Delta(T)$ has a clear BCS-like T-dependence. In addition, with increasing doping the two gaps cross over in magnitude, just as shown in Fig. 8.

The same principle applies to the ARPES data. The low-temperature gap will just be the larger of the two gaps. It is preferable, therefore, to plot the p-dependence of the normal-state gap just *above* $T_c$. Now the ARPES data is very clear. The leading-edge gap at 100K [61] falls roughly linearly with doping and disappears *abruptly* at critical doping as shown by the circles in Fig. 9. Furthermore, it is well known that normal-state quasiparticle (QP) lifetimes are strongly damped near ($\pi$,0) due to scattering from spin fluctuations with the result that the leading-edge normal-state QP peak is completely washed out in underdoped samples but not in overdoped samples. The data points on the $T_c(p)$ curve in Fig. 9 represent the $T_c$ values for all published ARPES spectra near ($\pi$,0) taken at 100K for Bi-2212 [32]. The open symbols denote spectra where the QP peak is washed out and the full symbols where it is recovered. The crossover occurs precisely at critical doping where the normal-state leading-edge gap disappears and, furthermore, is very sudden as shown by the two spectra [62,63] either side of the margin.

Finally in this section, we note that the normal-state pseudogap in earlier STM studies [16] was observed to persist well into the overdoped region before disappearing and, in particular, beyond critical doping. We believe this conclusion is not robust. This behaviour was found to extend to temperatures no higher than 90K i.e. no higher than $T_{c,max}$. In view of our above observation concerning surface inhomogeneity and the averaging effects of a mildly unstable tip, we presume that this persistent gap is just a vestige of the superconducting gap from regions of lower local doping.

*3.9 Zinc substitution*

We conclude our arguments with a brief discussion of the rather revealing effects on superconductivity and the pseudogap of Zn substitution for Cu. Zn is a non-magnetic pairbreaker that acts as a unitary scatterer in rapidly suppressing superconductivity [40,64] without altering the doping state [65,66]. (In passing we stress that Ni also acts as a unitary scatterer and suppresses $T_c$ as rapidly as Zn [67], as does Li [68]. Much has been made in the literature of a major distinction between the effects of these substituents now attributable simply to solubility effects). We note, firstly, the crucial observation that NMR Knight shift studies [65,69], heat capacity studies [18] and thermopower studies [70] all independently show that the pseudogap energy scale, $E_g$, remains unaltered by Zn substitution up to and well beyond levels sufficient to fully suppress superconductivity. In contrast, a mere 2% Zn in fully-oxygenated Y-123 reduces both $T_c$ and $\Delta_o$ by as much as 30% [71]. This in itself is sufficient to show that superconductivity and the pseudogap are independent and compete. Further confirmation comes from resistivity data which shows that T* is also independent of Zn concentration [38].

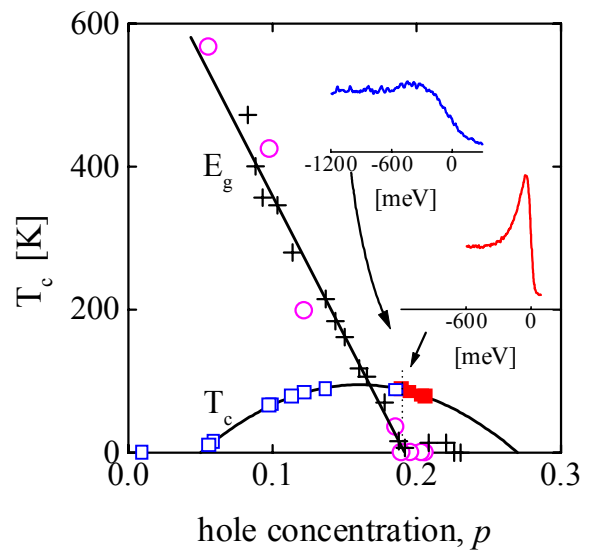

Fig. 9. The p-dependence of the leading-edge ARPES gap at 100K near the (0,$\pi$) zone boundary (○) and of $E_g$ found from heat capacity (+). $T_c$ values are given by squares: no QP peak at 100K (□); full QP peak at 100K (■). The two EDC spectra either side of critical doping are shown. Upper: Norman *et al.* ref. [62]; lower: Kim *et al.* ref. [63].

The substitution of Zn therefore allows one to "peel back" the superconductivity to progressively lower temperatures and observe the evolution of the pseudogap in a region of doping and temperature where the pseudogap correlations are usually concealed by superconductivity. For Y-123 with 2% Zn substitution slight underdoping almost completely suppresses $\Delta\gamma$ and $T_c$ exposing an unaffected normal-state pseudogap temperature dependence in both $\gamma$ and entropy, S [71]. This would appear to be a clear demonstration that superconductivity and the pseudogap are independent and competing. Zn substitution is particularly instructive in fully-oxygenated Y-123. In this compound one finds the fortuitous circumstance that full oxygenation corresponds pretty closely to critical doping ($YbBa_2Cu_3O_7$ with a smaller rare earth element lies beyond critical doping while $GdBa_2Cu_3O_7$ and other larger rare earth element 123 materials lie short of critical doping). Suppose then we concede that the pseudogap corresponds to precursor pairing, T* is the mean-field transition temperature and $T_c$ is the phase-coherence temperature. It is natural to associate the apparently universal doping state p=0.19 with the point at which T* breaks away from $T_c$ and phase fluctuations set in above $T_c$. The effect of Zn substitution is to rapidly reduce the condensate density (in fact, consistent with unitary scattering in a d-wave superconductor, substantially faster than the $T_c$ suppression [40]). The consequence, within the precursor pairing scenario, is that phase fluctuations would occur more readily and the point where T* breaks away from $T_c$ would move to a higher doping state, beyond p=0.19. This is not found experimentally. Indeed, the normal-state $\gamma(T)$ for fully-oxygenated $YBa_2(Cu_{1-x}Zn_x)_3O_7$ remains constant, independent of temperature and Zn content, for x=0, 1, 2, 3, 5 and 7% [71] in spite of a strongly reduced condensate density. Because these samples are critically doped they sit at the pseudogap threshold and we may conclude that there is no Zn-induced onset of the pseudogap (which would be signalled by a Zn-induced reduction in $\gamma$). Critical doping remains constant, independent of Zn substitution [18] and is not governed in any way by the magnitude of the condensate density. These results, therefore, strongly implicate *competing* correlations and not *precursor* correlations. This conclusion is echoed by a comparison of different cuprates. Critical doping appears to take a common value amongst the cuprates while the condensate density at critical doping may vary widely. $YBa_2Cu_3O_7$ and $Y_{0.8}Ca_{0.2}Ba_2Cu_3O_{6.81}$ both lie at critical doping while the condensate density of the former (measured by μSR) is nearly double that of the latter [17].

In fact, for $Y_{0.8}Ca_{0.2}Ba_2Cu_3O_{7-\delta}$, $La_{2-x}Sr_xCuO_4$ and $Bi_2Sr_2CaCu_2O_{8+\delta}$, the progressive suppression of both $T_c$ and condensate density are quantitatively reproduced as a function of both doping and Zn content within a model of unitary scattering for a d-wave order parameter (and with no adjustable parameters) under the assumption of a competing pseudogap which remains unaffected by Zn substitution and which vanishes at p=0.19 [64,40,72,73]. To give further weight to this conclusion we show in Fig. 10 the doping dependence of the critical fraction of Zn, $x_{crit}$, at which $T_c$ is suppressed to zero in $Y_{0.8}Ca_{0.2}Ba_2(Cu_{1-x}Zn_x)_3O_{7-\delta}$, plotted along with the entropy per Cu, $S(T_c)/k_B$ for the unsubstituted material, $Y_{0.8}Ca_{0.2}Ba_2Cu_3O_{7-\delta}$. Across the entire doping range $x_{crit}$ and $S(T_c)/k_B$ are nearly equal even though they change by as much as an order of magnitude. This shows that each Zn atom breaks one pair, as expected for unitarity limit potential scat-tering which predicts [74] $x_{crit} \approx 1.3\, N_o\, \Delta_o \sim \rho_{pair}(0)$, the pair density at T=0. We may conclude that the strongly reduced S and $\gamma$ at $T_c$ in underdoped samples does indeed reflect a sharply reduced $\rho_{pair}(0)$ consistent with a loss of spectral weight due to a competing correlation. $\rho_{pair}(0)$ would not be affected if $S(T_c)$, $\gamma(T_c)$ and $T_c$ were all reduced due to phase fluctuations (a high temperature effect) as in precursor pairing models. Most significantly, Fig. 10 implies conventional BCS-like behaviour (albeit strong coupling) across the entire superconducting phase diagram.

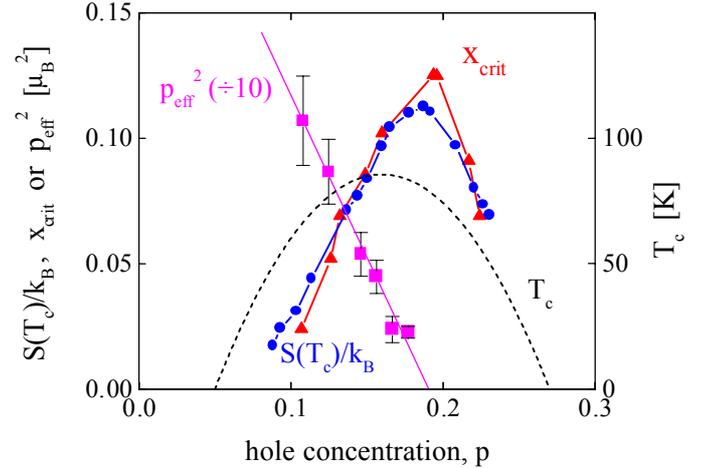

Fig. 10. The doping dependence of the critical zinc concentration, $x_{crit}$, for suppressing superconductivity in $Y_{0.8}Ca_{0.2}Ba_2(Cu_{1-x}Zn_x)_3O_{7-\delta}$ (▲) and of the entropy at $T_c$ per copper atom in $Y_{0.8}Ca_{0.2}Ba_2Cu_3O_{7-\delta}$ (●). Across the entire doping range $x_{crit} = S(T_c)/k_B$ per copper revealing the essentially BCS-like behaviour at all doping. Also shown (■) is the square of the effective Zn-induced moment for $YBa_2(Cu_{1-x}Zn_x)_3O_{7-\delta}$ obtained from the Curie term in the bulk static susceptibility [76] indicating that AF correlations disappear quite suddenly at critical doping.

Finally, we consider the role of Zn substitution where there is a background of short-range AF correlations. A common view is that, though Zn is non-magnetic, its substitution into an AF-ordered environment will induce a local moment on the four neighbouring coppers due to the uncompensated spins about the spin vacancy. $^{89}$Y-NMR studies reveal a satellite resonance arising from an yttrium atom immediately adjacent to the Zn atom and the Knight shift of this satellite has a pure Curie temperature dependence [75,67]. This suggests that the pseudogap is locally suppressed immediately adjacent to the Zn atom and the local susceptibility is dominated by the local moment. Using macroscopic magnetic susceptibility measurements, Alloul et al. [76] have measured the oxygen concentration dependence of the Curie constant for the local moment induced by Zn in $YBa_2(Cu_{0.96}Zn_{0.04})_3O_{7-\delta}$. We have replotted this as a function of hole concentration in Fig. 10. The induced moment can be seen to fall progressively towards zero at critical doping indicating, again, that the AF background disappears abruptly at critical doping just where the pseudogap disappears. This further confirms that the pseudogap is linked closely to short-range, though relatively long-lived, AF correlations and does not arise from incoherent pairing correlations.

The very recent study by Bobroff et al. [68] using $^7$Li NMR revealed a Kondo-like $1/(T+\theta)$ temperature dependence of the Knight shift $^7K_s(T)$ persisting into the overdoped regime in samples of $Y_{0.8}Ca_{0.2}Ba_2Cu_3O_{7-\delta}$ which led them to infer an induced moment even in the overdoped regime. However, such a $1/(T+\theta)$ dependence occurs in pure (Zn- or Li-free) samples in the $^{89}$Y NMR Knight shift for $Y_{0.8}Ca_{0.2}Ba_2Cu_3O_{7-\delta}$ [77] and also in S/T for $Bi_2Sr_2CaCu_2O_{8+\delta}$ [78] and is, therefore, intrinsic.

## 4. Conclusions

The above observations make it absolutely clear that all of the above-mentioned spectroscopies are uniformly consistent in showing that the pseudogap energy scale vanishes at critical doping as shown in scenario (a) in Fig. 1. To underscore this conclusion we collect together in Fig. 11 all of the data for $E_g$ quantified in the above discussion. This includes DC susceptibility [79], heat capacity [18], Knight shift [23], ARPES [61] and resistivity [19]. This data, presented as reported in the literature without modification, is remarkably consistent in showing an energy scale that has the magnitude of J in the undoped insulator and falls uniformly and abruptly to zero at the same critical doping in each case. The pseudogap is seen to be independent of the superconductivity (which persists to a much higher doping state p=0.27). Moreover, Raman, spin-lattice relaxation rate, ARPES and Zn-substitution effects all uniformly show that the pseudogap is intimately connected with (though not equivalent to) short-range AF correlations which disappear at the same critical doping state. Inelastic neutron scattering (INS) studies on Y-123 confirm this [80] though, in the case of La-214, incommensurate INS peaks around $(\pi/2,\pi/2)$ appear to persist out to higher doping (p=0.25) [81]. This could be due to the presence of stripes in the latter compound which may not be present in other cuprates. Such a view is supported by the anomalous thermopower of La-214 [41].

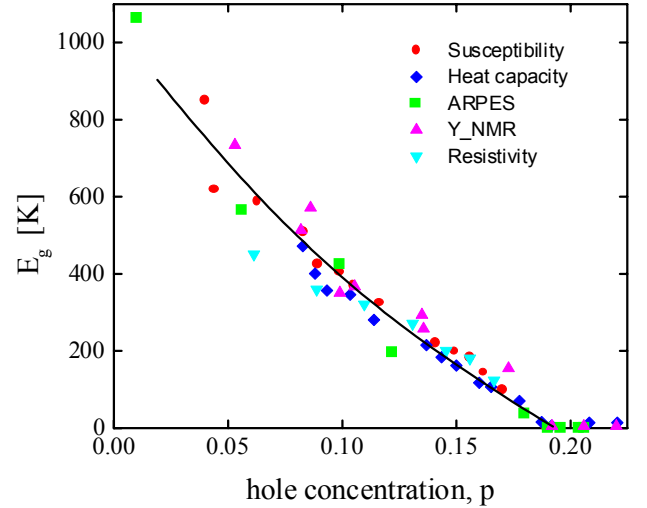

Fig. 11. The p-dependence of the pseudogap energy, $E_g$, determined from susceptibility [79], heat capacity [18], ARPES [61], $^{89}$Y-NMR [23] and resistivity [19].

The energy scale $E_g(p)$ is clearly linked with the crossover temperature $T_{cr}(p)$ discussed by Pines [82,83] where the AF correlation length $\xi_{AF}(p,T)$ becomes less than 2a. While it is common in the field to refer to *underdoped* and *overdoped* samples with respect to maximal $T_c$ ($p<p_{opt}=0.16$ and $p>p_{opt}=0.16$, respectively) we concur with Pines [83] that the physics is better represented by the notion of *magnetic underdoping* and *magnetic overdoping*. Based on the present summary (and especially with reference to Figs. 4 and 5) we may say that magnetically underdo-ped samples, with $p_o<p_{crit}=0.19$, exhibit *weak superconductivity* with a sharply reduced condensation energy while magnetically overdoped samples, with $p>p_{crit}=0.19$, exhibit *strong superconductivity* [73]. The physics is such that optimal doping, where $T_c$ maximises, has no special meaning and indeed, with increasing impurity scattering, $p_{opt} \to p_{crit}$ as $T_c \to 0$ [64].

A critical interpretation of the data thus shows that scenario (b) in Fig. 1, with $T^* \to T_c$, is not sustainable. Along with this it follows that the pseudogap cannot be precursor pairing which leads to the coherent superconducting state below $T_c$. By all means it could be a *competing* precursor pairing of mechanism different from that leading to the superconducting state but the fact that the energy scale rises to J at low doping, and the pseuodogap's intimate relation to short-range AF correlations, suggests this is rather unlikely. It is hoped that these observations on the phenomenology of the pseudogap will help clear the way to proceed more rapidly to a consensus on the fundamental mechanisms involved in high-$T_c$ superconductivity.

**Acknowledgements**

Thanks are due to Dr. C. Bernhard, Dr. G.V.M. Williams and Dr. J. R. Cooper for ongoing discussion on the issues raised herein. This work was carried out under a Visiting Fellowship, Trinity College, Cambridge (JLT).

**References**

[1] T. Timusk and B. Statt, Rep. Prog. Phys. **62**, 61 (1999).
[2] H. Alloul, T. Ohno and P. Mendels, Phys. Rev. Lett., **63**, 1700 (1989).
[3] H. Yasuoka, T. Imai and T. Shimizu in *Strong Correlations and Superconductivity* ed. by H. Fukuyama, S. Maekawa and A.P. Malozemoff, Springer Series in Solid State Sciences, Vol. 89 (Springer, Berlin, 1989), p. 254.
[4] J.L. Tallon *et al.*, Physica C **235-240**, 1821 (1994).
[5] V.J. Emery and S.A. Kivelson, Nature (London) **374**, 434 (1995).
[6] T.M. Rice, Phys. World **12**, (12) 55 (1999).
[7] J.L. Tallon *et al.*, phys. stat. solidi (b) **215**, 531 (1999).
[8] N. Nagaosa and P.A. Lee, Phys. Rev. Lett., **64**, 2450 (1990).
[9] R.S. Markiewicz, Phys. Rev. B **56**, 9091 (1997).
[10] B. Bucher *et al.*, Phys. Rev. Lett., **70**, 2012 (1993).
[11] T. Ito, K. Takenaka and S. Uchida, Phys. Rev. Lett., **70**, 3995 (1993).
[12] B. Batlogg *et al.*, Physica C **235-240**, 130 (1994).
[13] H.Y. Hwang *et al.*, Phys. Rev. Lett., **72**, 2636 (1994).
[14] J.W. Loram *et al.*, J. Supercon. **7**, 243 (1994).
[15] H. Ding *et al.*, Nature (London) **382**, 51 (1996).
[16] Ch. Renner *et al.*, Phys. Rev. Lett., **80**, 149 (1998).
[17] J.L. Tallon *et al.*, Phys. Rev. Lett., **74**, 1008 (1995).
[18] J.W. Loram *et al.*, J. Phys. Chem. Solids **59**, 2091 (1998).
[19] B. Wuyts, V.V. Moshchalkov and Y. Bruynseraede, Phys. Rev. B **53**, 9418 (1996).
[20] J. Schmalian, D. Pines and B. Stojkovic, Phys. Rev. B **60**, 667 (1999).
[21] M.R. Presland *et al.*, Physica **165C**, 391 (1991).
[22] G.V.M. Williams *et al.*, Phys. Rev. Lett., **80**, 377 (1998).
[23] J.L. Tallon *et al.*, J. Phys. Chem. Solids **59**, 2145 (1998).
[24] J.W. Loram *et al.*, Advances in Superconductivity VII (Springer, Tokyo, 1995) p. 75.
[25] J.W. Loram *et al.*, Proc. 6[th] Internat. Conf. Materials and Mechanisms of Superconductivity, Physica C (in press)
[26] F. Raffa *et al.*, Phys. Rev. Lett., **81**, 5912 (1998).
[27] A.J. Millis, H. Monien and D. Pines, Phys. Rev. B **42**, 167 (1990).
[28] G.V.M. Williams, J.L. Tallon and J.W. Loram, Phys. Rev. B **58**, 15053 (1998).
[29] J.L. Tallon, G.V.M. Williams and D.J. Pringle, cond-mat/9902180.
[30] S. Ohsugi *et al.*, J. Phys. Soc. Japan **60**, 2351 (1991).
[31] J.W. Loram *et al.*, *10[th] Anniversary HTS Workshop* (World Scientific, 1996), p. 341.
[32] J.L. Tallon, Advances in Superconductivity XII (Springer, Tokyo,1999) (in press); cond-mat/9911422.
[33] N.E. Hussey *et al.*, Phys. Rev., B **56**, 11423 (1997).
[34] N.Y. Chen *et al.*, Phys. Rev., B **50**, 16125 (1994).
[35] Z. Konstantinovic, Z.Z. Li and H. Raffy, Physica C (in press).
[36] W.E. Lawrence and S. Doniach, in Proceedings of the 12[th] International Conference on Low Temperature Physics, Kyoto, Japan, 1970, ed. by E. Kandi (Keiyakol, Tokyo, 1970), p. 361.
[37] D.J.C. Walker, A.P. Mackenzie and J.R. Cooper, Phys. Rev. B **51**, 15653 (1995).
[38] K. Mizuhashi *et al.*, Phys. Rev. B **52**, R3884 (1995).
[39] G. Boebinger *et al.*, Phys. Rev. Lett., **77**, 5417 (1996).
[40] C. Bernhard *et al.*, Phys. Rev. Lett., **77**, 2304 (1996).
[41] J.R. Cooper and J.W. Loram, J. Phys. IV (France) **10**, 213 (2000).
[42] C. Bernhard *et al.*, Phys. Rev., B **59**, R6631 (1999).
[43] R. Henn *et al.*, Phys. Rev., B **56**, 6295 (1997).
[44] S. Uchida, K. Tamasaku and S. Tajima, Phys. Rev., B **53**, 14558 (1996).
[45] T. Xiang and J.M. Wheatley, Phys. Rev. Lett., **77**, 4632 (1996).
[46] D. van der Marel, Phys. Rev. B **60**, 765 (1999).
[47] D. van der Marel *et al.*, cond-mat/0003157.
[48] N. Miyakawa *et al.*, Phys. Rev. Lett., **83**, 1018 (1999).
[49] D. Munzar, Sol. State Comm. **112**, 365 (1999).
[50] M. Grueninger *et al.*, Phys. Rev. Lett., **84**, 1575 (2000).
[51] C. Bernhard (private communication).
[52] C. Kendziora and A. Rosenberg, Phys. Rev., B **52**, R9867 (1995).

[53] J.G. Naeini *et al.*, Can. J. Phys (in press); cond-mat/9909342.
[54] M. Rübhausen *et al.*, Phys. Rev. Lett., **82**, 5349 (1999).
[55] G. Blumberg *et al.*, Science, **278**, 1427 (1997).
[56] K.C. Hewitt *et al.*, Phys. Rev., B **60**, R9943 (1999).
[57] J.W. Loram *et al.*, Phys. Rev. Lett., **71**, 1740 (1993).
[58] V. Madhavan *et al.*, Bull. Amer. Phys. Soc., **45** (1), 416 (2000).
[59] J.L. Tallon and J.W. Loram, cond-mat/.
[60] V.M. Krasnov *et al.*, cond-mat/0002172
[61] R.B. Laughlin, Phys. Rev. Lett., **79**, 1726 (1997).
[62] M.R. Norman *et al.*, Phys. Rev. Lett., **79**, 3506 (1997).
[63] C. Kim *et al.*, Phys. Rev. Lett., **80**, 4245 (1998).
[64] J.L. Tallon *et al.*, Phys. Rev. Lett., **79**, 5294 (1997).
[65] H. Alloul *et al.*, Phys. Rev. Lett., **67**, 3140 (1991).
[66] G.V.M. Williams and J.L. Tallon, Phys. Rev. B **59**, 3911 (1999).
[67] G.V.M. Williams, J.L. Tallon and R. Dupree, Phys. Rev. B **61**, 4319 (2000).
[68] J. Bobroff *et al.*, Phys. Rev. Lett., **83**, 4381 (1999).
[69] G.V.M. Williams *et al.*, Phys. Rev. B **51**, 16503 (1995).
[70] J.L. Tallon *et al.*, Phys. Rev. Lett., **75**, 4114 (1995).
[71] J.W. Loram *et al.*, Physica C **235-240**, 134 (1994).
[72] G.V.M. Williams, E.M. Haines and J.L. Tallon, Phys. Rev. B **57**, 5146 (1998).
[73] J.L. Tallon, Phys. Rev. B **58**, R5956 (1998).
[74] Y. Sun and K. Maki, Phys. Rev. B **51**, 6059 (1995).
[75] A.V. Mahajan *et al.*, Phys. Rev. Lett., **72**, 3100 (1994).
[76] H. Alloul *et al.*, High Temperature Superconductivity, ed. by S.E. Barnes et al., (AIP Conference Proceedings, 1999, vol. 483) p. 161.
[77] G.V.M. Williams *et al.*, Phys. Rev. B **57**, 8696 (1998).
[78] J.W. Loram *et al.*, Physica C (in press).
[79] J.R. Cooper and J.W. Loram, J. Phys. I France **6**, 2237 (1996).
[80] P. Bourges, in *Gap Symmetry and Fluctuations in High Temperature Superconductors* ed. by J. Bok, G. Deutscher, D. Pavuna and S.A. Wolf (Plenum, 1998).
[81] K. Yamada *et al.*, Phys. Rev. B **57**, 6165 (1998).
[82] D. Pines, Physica C **282-287**, 273 (1997).
[83] D. Pines, cond-mat/0002281.